# Self-Assembly of Carbon-Nanotube-Based Single Electron Memories**

Laëtitia Marty, Anne-Marie Bonnot, Aurore Bonhomme, Antonio Iaia, Cécile Naud,

Emmanuel André and Vincent Bouchiat *

[*] Dr. L. Marty, Dr. A.-M. Bonnot, A. Iaia, Dr. C. Naud,

Laboratoire d'Etudes des Propriétés Electroniques des Solides

CNRS-UJF, BP 166, Grenoble Cedex 9, France.

A. Bonhomme , E. André, Dr. V. Bouchiat

Centre de Recherches sur les Très Basses Températures,

CNRS-UJF, BP 166, Grenoble Cedex 9, France.

FAX : (+33) 476 87 50 60

Email : bouchiat@grenoble.cnrs.fr

[**] This research is supported by the French Ministry of Research through the Action Concertée Nanosciences program and by the Region Rhône-Alpes. We gratefully acknowledge help from J. Chaussy with the parylene deposition and B. Pannetier for his constant support during this work.

# Abstract

We demonstrate the wafer-scale integration of single electron memories based on carbon nanotube field effect transistors (CNFETs) using a process based entirely on self assembly. First, a "dry" self assembly step based on chemical vapor deposition (CVD) allows the growth and connection of CNFETs. Next, a "wet" self-assembly step is used to attach a single 30 nm-diameter gold bead in the nanotube vicinity *via* chemical functionalization. The bead is used as the memory storage node while the CNFET operated in the subthreshold regime acts as an electrometer exhibiting exponential gain. Below 60 K, the transfer characteristics of gold-CNFETs show highly reproducible hysteretic steps. Evaluation of the capacitance confirms that these current steps originate from the controlled storage of single electrons with a retention time that exceeds 550 s at 4 K.

# 1. Introduction

Single electron devices could be a possible breakthrough in tomorrow's nanoelectronics [1]. Indeed the fundamental quantization of the electric charge could be used to code and process digital information. Reduced power consumption, and large scale integration are among the many advantages. The most promising single electron device for a broad range of applications is the single electron memory [2]. The single electron memory can be seen as the ultimate miniaturization of the flash memory for which Coulomb blockade can lead to multilevel coding functionality [3]. In this device, Coulomb blockade introduces additional rules that allow an external gate to control the exact number of excess electrons in the storage node. This node consists of a small capacitance electrode (the so-called "island") which acts as a charge trap. When placed in the Coulomb blockade regime the island can exchange electrons one by one via tunnelling with a charge reservoir.. To allow the charge readout, the island is capacitively coupled to a charge sensitive amplifier device (the so-called "electrometer") that must have sub-electron charge sensitivity [4]. The electrometer can be implemented either by a single electron transistor (SET), as in the original experiments, [2] or by a highly sensitive field effect transistor (FET) [3]. Many kinds of electrometers have been implemented so far, that involve metallic [2], semiconducting [5] and more recently single walled [6, 7] or multi walled [8] carbon nanotube as the active channel.

Up to date, fabricating a single electron memory has involved at least one electron beam lithography and mask alignment steps thus forbidding low-cost, large scale integration. In this paper, we present a method based on self-assembly that allows the successful parallel integration of single electron memories for which the island is implemented by a gold nanocrystal and the electrometer by a Carbon Nanotube FET (CNFET) (Figures 1 and 2).

## 2. Results

### 2.1 CNFET integration and functionalization

The CVD growth of carbon nanotubes using catalyst covered electrodes has proven to be a powerful method to achieve the full integration of CNFET devices [9]. In our process, we perform this template directed growth using a hot-filament-assisted CVD method .[10] The growth parameters are adjusted to reach the limit of an individual carbon nanotube or bundle connection per electrode pair [11]. This technique allows the simultaneous self-assembly and electrical connection during the nanotube synthesis. Both unipolar and ambipolar CNFETs [12] are obtained with a yield that exceeds 50% on the 9000 electrode pairs. As confirmed by TEM and AFM measurements (see supplementary material), our synthesis method leads to SWNTs having relatively large diameters (that often exceed 1.5nm) with a probable high occurrence of double walled carbon nanotubes. Semiconducting nanotubes will then have a band gap in the range 0.4-0.6eV [13] which is almost half of what is currently seen in CVD prepared SWNTs. Because of their relatively low band gap, these nanotubes like is seen in double walled carbon nanotubes [14] are quite sensitive to electrostatic doping, for both positive and negative gate voltages yielding to a large occurrence of ambipolar CNFETs [15]. In our batches, more than half of the connected circuits exhibits an ambipolar field effect, which is significantly higher compared to the other CVD methods.

For the deposition of the colloidal gold beads we proceed to a standard silanization of the silica substrate. This silanization process induces a net positive charge on the substrate by grafting amine groups on the silicon oxide of the wafer surface. Such a surface functionalization is known [16] to lead to well pronounced charge transfer that enhances the *p* type behaviour in the already p-doped CNFETs, as shown in Figure 1.

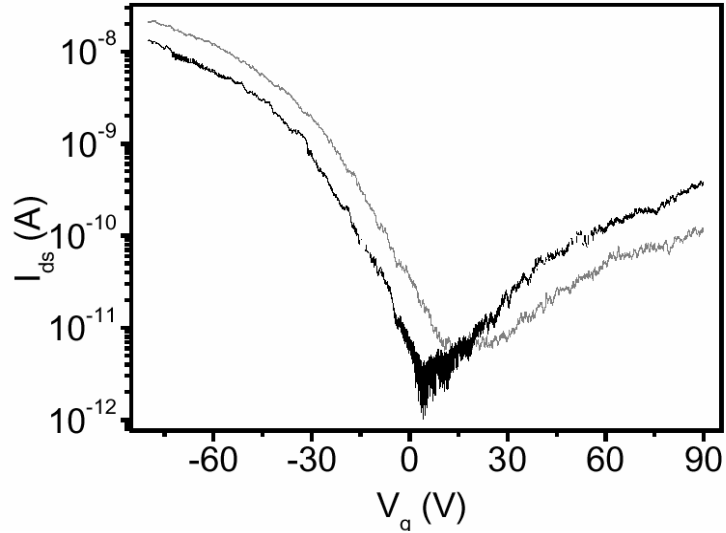

Figure 1 : Current versus gate voltage of a typical, 1-μm long self-assembled CNFET before (black curve) and after (gray curve) silanization of the silica substrate measured at low temperature (4K). The drain-source voltage is 2V while the gate voltage was applied using silicon backgating. The positive charging of the substrate induced by the surface functionalization induces a charge transfer seen as an effective *p* doping of the CNFET. The same effect is seen at room temperature but is less pronounced.

Compared to the field effect measured before silanization, threshold voltages are shifted towards negative gate voltages. Moreover the saturated "On current" on the *p* side (negative gate voltages) increases by an amount of 100% at 4.2 K, while the current on the *n* side decreases by the same amount. Once the CNFETs fabricated by the "dry" self assembly step are fully characterized, we proceed to the second self assembly step which consists in "wet" coupling a single gold nanoparticle to each CNFET. A relatively large diameter (30nm) was chosen for reproducibility and for increased electrostatic coupling. By varying the colloid concentration we can obtain a bead density of about 1 $\mu m^{-2}$ which leads on average to a single bead coupling per CNFET (See Figure 2). Even with such a low surface coverage, the nanotube/colloid distance was found to be around 10 nm in most cases. Grafting of aminosilane in our process lead to a relatively thick capping layer on the silica and on the nanotube. The good adhesion probability of the bead near the nanotube is attributed to the combined effects of the surface roughness created near the nanotube and to the presence of amine groups even on the nanotube.

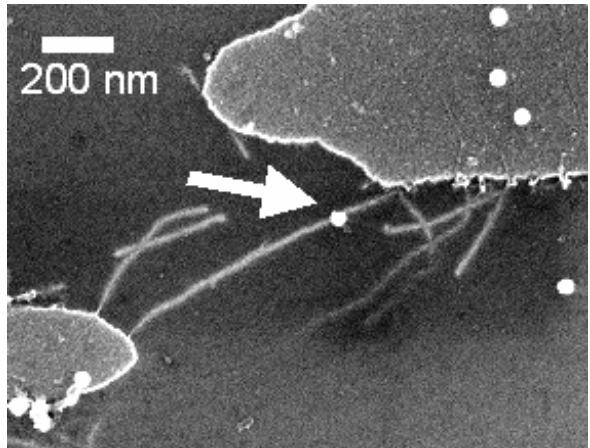

Figure 2: Scanning Electron micrograph showing a typical self assembled carbon nanotube based single electron memory. It is shown after the two self-assembly steps but before parylene encapsulation and top gating. The nanotubes are lying on the silanized silica substrate and a single 30-nm diameter gold nanocrystal (arrow) is deposited.

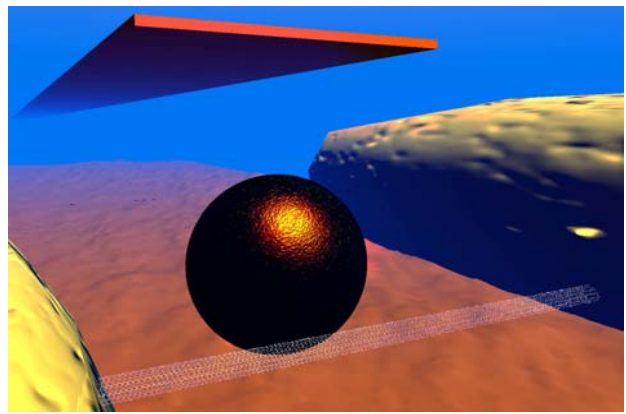

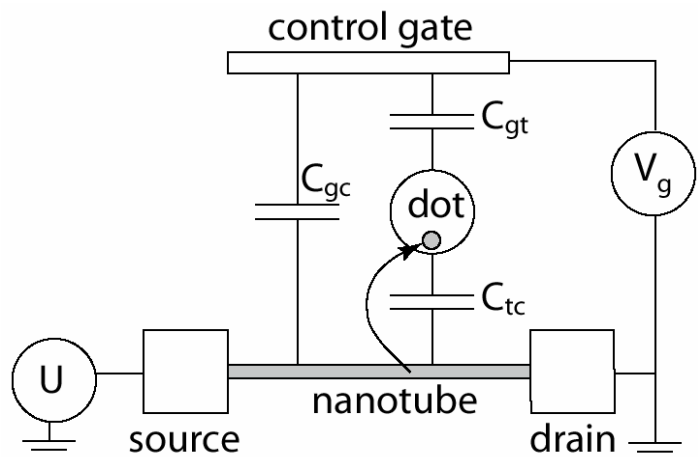

Figure 3: Top: Artist view of the nanotube-based memory device. The nanotube electrometer is self-assembled via the HFCVD technique between metallic electrodes, (note that relative heights of the nanotube tha electrodes and the gold bead is at scale) . The colloidal gold nanocrystal is self-assembled by a physico-chemical process. A top gate is realized above the gap between the electrodes. Bottom: Equivalent electrical circuit showing the different capacitances involved in the gate coupling of the nanotube in the memory device. Charge transfers occur between the nanotube and the gold nanocrystal as shown by the arrow.

## 2.2 Electrical characterization of the memory effect

For characterization, the memory device is encapsulated in a 100-nm-thick parylene C dielectric layer, a material well suited for deposition on top of molecular devices [17]. Finally a gate electrode was deposited on top to increase the gate coupling. No nanotube-gate leakage was measured using the top gate, thanks to the parylene that allows a highly conformal deposition of a dielectric film.

Below 120 K the field effect exhibits steps (Figs. 4,5) that become hysteretic below 60 K. These steps are direct measurements of the charge transfer that occurs between the CNFET and the gold bead. This is confirmed by the fact that no steps are observed for transistors without the gold bead attachment (Figure 4 a) inset).

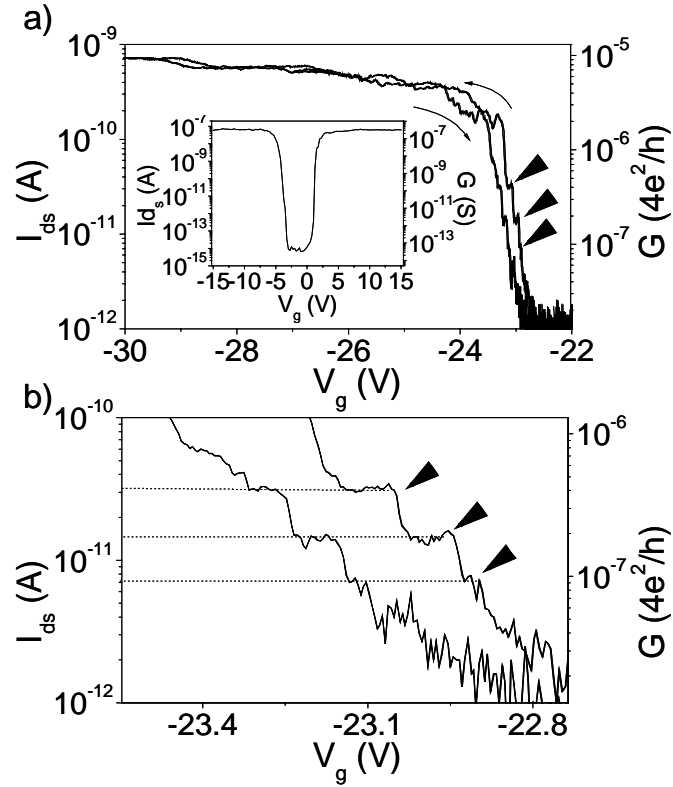

Figure 4: a) CNFET current as versus gate voltage at 2.2 K for a typical device (sample A). The curve is plotted at $V_{ds}$=500 mV for opposite gate sweeps of ±0.5 V.s$^{-1}$. Inset: Transfer characteristic of a control CNFET sample from the same batch but without gold functionalization for which no current step is observed. The drain-source current was measured at $T$=6 K and for drain-source voltage $V_{ds}$= 500 mV. The gate voltage was applied on both the top and the back gates with a sweep rate of -0.1 V.s$^{-1}$. The characteristic exhibits symmetric "*p*" and "*n*" type branch showing an ambipolar behavior. b) Zoom on the above curve in the subthreshold regime. In this semilog scale the nanotube current exhibits regularly spaced steps in both X and Y directions, signature of a constant amount of charge transfer between the nanotube and the gold bead.

For a 30 nm diameter gold bead, the self capacitance is $C_{self} = 4\pi\varepsilon_0\varepsilon_r r = 5,3$ aF (with $r=15$ nm the sphere radius and $\varepsilon_r = 3,2$ the parylene C dielectric permittivity). The charging energy is $E_c = \frac{e^2}{2C_{self}} = 15$ meV equivalent to a temperature of 180 K. At temperatures well bellow this limit, Coulomb blockade dominates electron transport, so single electrons transit between the channel (*i.e.* the CNT) and the floating dot (*i.e.* the gold bead) through a thick junction (either multiple tunnel junctions or a Fowler-Nordheim field emission barrier) [18]. The transfer occurs then at regularly spaced gate voltages separated by $\Delta V_w$ (Figure 5) where $e\Delta V_w$ corresponds to the voltage source work necessary to add or remove charge quanta from the bead. The charge stored in the bead induces a local electrostatic field on the nanotube which in turn modifies its conductivity. The CNFET thus acts as an electrometer probing the charge present on the bead. This phenomenon is reminiscent to what occurs in CNFET-based electrochemical transducers [19] for which metal particles can be coupled to the CNFET channel to make the sensing properties chemically specific [20]. In the subthreshold region, the CNFET transconductance response is exponential: $\Delta I_{ds} = \frac{1}{S(T)}\exp\left(\frac{C_{gc}\Delta V_g}{e}\right)$, where $S(T)$ is the temperature dependent subthreshold slope. Therefore any electrostatic charge modification in the bead will be exponentially amplified. Indeed, this amplification originates from the fact that the current through the nanotube is established via quantum electron hopping [12,21] which can be affected by the local electrostatic environment [22,23]. This exponential behaviour in the substhreshold regime allows the observation of regularly spaced current steps when plotted in *log($I_{ds}$)* Vs $V_g$ (see Figures 4 and 5), contrary to its operation in the saturation region (between –30 V and –24 V on Figure 4 ) for which the charge sensitivity rapidly drops.

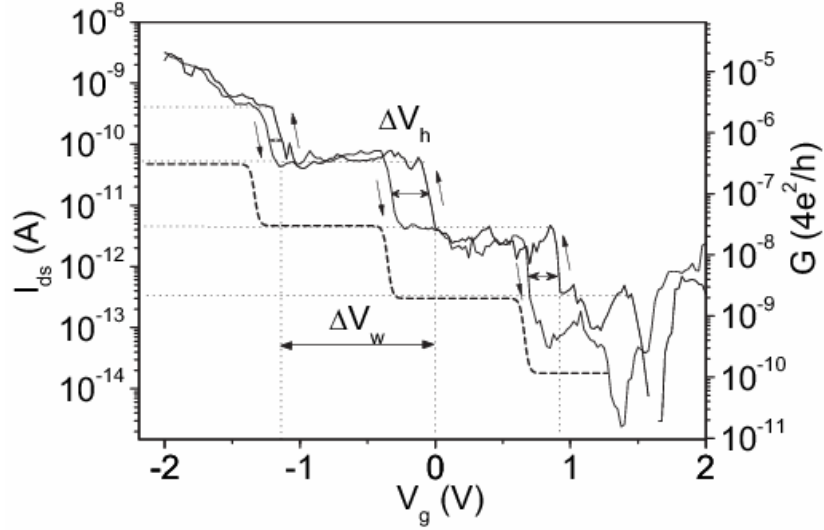

Figure 5: Memory operation of a second gold-functionalized CNFET device (sample B) probed in the subthreshold regime at $T$=2 K with a low sweep rate 10 mV.s$^{-1}$. Dotted line: a fit (shifted for clarity) obtained from the from the formula 1 of the Coulomb staircase of an electron box. The charging energy of the fit is 80K. Time retention of an electron in the gold beads exceeds 550 seconds.

A precise analysis of the memory operation is provided for sample B (Figure 5). In this device, the CNFET current steps occur at regularly spaced gate voltages $\Delta V_w=0.93\pm 0.07$ V and, each current step is of about 1 decade high. This is sensitive enough for detecting even the inherently low channel currents circulating in the subthreshold regime. As expected, this charge trapping is not symmetrically reversible: the charging/discharging of the trap occurs at gate voltage separated by $\Delta V_h$. The induced hysteresis depends on the temperature and on the gate voltage sweep rate. Indeed charge retention in the Coulomb blockade regime is easily affected by thermal activation [23]. At low temperatures ($T \sim 2$ K), the hysteresis $\Delta V_h$ is relatively constant even for a low sweep rate (Figure 5). Its value $\Delta V_h=0.18\pm 0.10$ V has been measured between two corresponding steps on the gate sweeps. These steps correspond to the trapping an detrapping of the same excess charge on the particle and are separated by a time $\tau$ related to the sweep rate and cycle amplitude. We have measured the current at a very slow sweep rate of ±10 mV.s$^{-1}$, and for an hysteresis cycle down to $V_g$ = -2 V, which implies $\tau$ is longer than 550 s. This time corresponds to the retention time of the charge before it is transferred away from the particle. This shows that the operation time of our memory can be longer than 550 s.

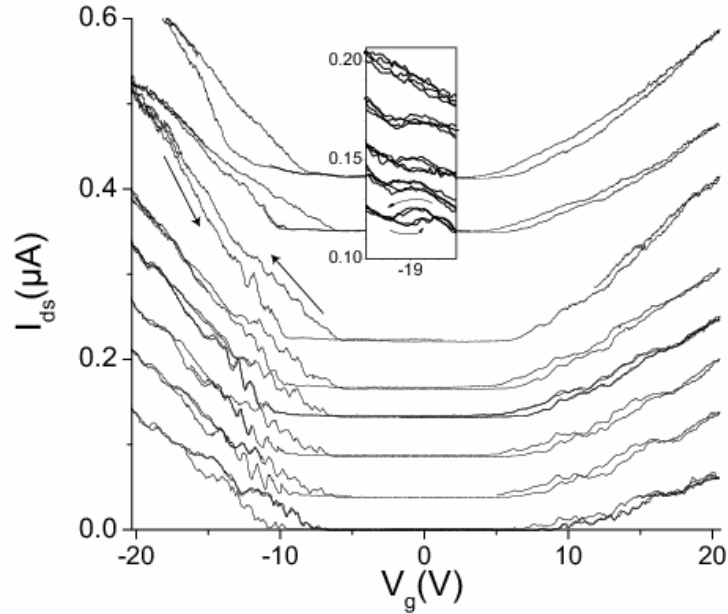

Figure 6: a) Temperature dependence of the transfer characteristics (Ids-Vg) for the memory (sample A). Curves are measured at $V_{ds}$=300 mV for high speed gate sweeps of 5V.s$^{-1}$. Curves are shifted for sake of clarity. From bottom to top, the temperatures are 15, 52, 60, 70, 90, 130, 160 and 225K. Current measurements at 160 and 225 K have been vertically rescaled by a factor ½. Inset : Zoomed data around Vg=-19V for the four lowest temperatures.

In order to provide another evidence to the fact that the gold bead is involved in the charge transfer, we have measured the temperature dependence of the memory effect in sample A. For that experiment we have swept the gate at a large amplitude, thus leading to a threshold voltage that is significantly shifted from those shown in Figure 4 on the same sample. Sharp steps due to discrete charge transfers vanish gradually while the temperature reaches the 100K-150K region. The inset shows a detail of a step showing the hysteresis disappearance that occurs at lower temperature (60K) than the total thermal smearing of the charge step (above 90K).

## 3. Operation of the memory in the single electron limit

In the low temperature regime $k_B T / E_c \ll 1$, the step depicting a single electron charging event for a single electron box can be written as [24]:

$$\langle n \rangle = m + \frac{1}{2}\left\{1 + \tanh\left[-\frac{E_c}{k_B T}\left(m + 1/2 - \frac{C_{gt} V_g}{e}\right)\right]\right\}. \quad \text{[Eq. (1)]}$$

This results is valid up to $E_c / k_B T \sim 0.2$. [25].

From that equation we deduce that the maximum slope of the Coulomb staircase is given at the point for which the argument of the hyperbolic tangent cancels to zero :

$$\left[\frac{\partial \langle n \rangle}{\partial (C_{gt} V_g / e)}\right]_{\max} = \frac{E_c}{2 k_B T}. \quad \text{[Eq. (2)]}.$$

One sees from that result that a noticeable thermal smearing of the Coulomb staircases occurs rapidly when the ratio $k_B T / E_c$ exceeds 0.1 [24]. A fit of the Coulomb staircase at the lowest temperature (2K) taking into account the exponential gain of the nanotube electrometer is shown in Figure 5. the fitted charging energy (80K) underestimate the predicted charging energy of the bead (180K). Several reasons can lead to such a discrepancy but the major uncertainty could come from spurious heating effect that is likely to occur in a cryogenic prober at such low temperatures.

To further investigate the thermal smearing of the discrete charge steps we have measured the temperature dependence of the transfer characteristic of the nanotube Ids(Vg) for sample A (Figure 6). As expected the sharp steps due to discrete charge transfers vanish gradually while the temperature reaches the 100K-150K region. Charge transfer involving surface contaminants would have persisted up to much higher temperature [26].

A closer analysis of the hysteresis (Figure 3) can give quantitative estimate of the transferred charge. The step parameters $\Delta V_w$ and $\Delta V_h$ characterize the charge increase and directly depend on the different coupling capacitances in the circuit and on the number $n$ of electrons transferred to or from the gold bead. This allows us to confirm the device is operated in the single electron limit by evaluating $n$ from the equation system [26]:

$$\begin{cases} \Delta V_w = \dfrac{ne}{C_{gt}} \\ \Delta V_h = \dfrac{neC_{tc}}{\left(C_{tc} + C_{gc}\right)C_{tot}} \end{cases} \qquad \text{[Eq. (3)]}$$

where $C_{tot}$ is the total gate-device capacitance given by:

$$C_{tot} = C_{gc} + \dfrac{C_{tc}C_{gt}}{C_{tc} + C_{gt}} \qquad \text{[Eq. (4)]}$$

Considering the presence of Schottky barriers at the nanotube-electrode contacts [27], one should expect an influence of the gold bead on the nanotube conductivity only if located near the metal/nanotube interface. However we observe the charge steps independently on the bead position along the nanotube. Additionally, the current does not vary between two discrete steps and does not exhibit the characteristic slope of the field effect. At higher temperatures (around 10 K), the bead screening becomes less efficient and the staircases start to bend and rapidly become equal to the average subthreshold slope. This suggests that the sensing part of the CNFET is very temperature dependent and that at low temperatures, the sensing area can be efficiently screened by the gold bead. The disorder along the nanotube length induces quantum dots connected in series [12] through which the electrons transit according to Coulomb blockade rules. Therefore the gold bead charge controls the electron percolation through the CNFET via an electrostatic action on a single quantum dot barrier. This specific dot acts as a filter for the whole device. It is then reasonable to assume that $C_{tc} \gg C_{gt}, C_{gc}$, which allows to simplify the equations according: $C_{tot} = C_{gc} + C_{gt}$, and $\Delta V_h = \dfrac{ne}{C_{gc} + C_{gt}}$.

These equations give the formula for *n*:

$$n = \dfrac{\Delta V_h C_{gc}}{e\left(1 - \dfrac{\Delta V_h}{\Delta V_w}\right)}. \qquad \text{[Eq. (5)]}$$

$C_{gc}$ is the capacitance between the nanotube intramolecular quantum dot and the gate. This parameter depends on the quantum dot size, which is here difficult to estimate without precise control experiments such as scanning gate microscopy [26] or Coulomb blockade measurements. Devices from other batches have been measured at very low temperatures (70 mK) which allowed us to determine the intramolecular quantum dot sizes from their charging energy and excited states [see supplementary material]. We thus assume the dot size *L* in this memory device is equal to the mean size obtained from other devices: *L*≈64±50 nm (such an assumption introduces a large uncertainty). $C_{gc}$ is then calculated using the geometric configuration of the gate:

$$C_{gc} = L \dfrac{4\pi\varepsilon_0\varepsilon_r}{2\ln\left(\dfrac{h + \sqrt{h^2 - a^2}}{a}\right)} \qquad \text{[Eq. (6)]}$$

where *h* is the dielectric thickness and *a* the nanotube radius. For sample B, the post measurement scanning electron microscopy observation of the device revealed a misalignment of the top gate inducing a very weak top-gating, therefore we only consider the back gate in the following calculations: *h* is then equal to 1 µm SiO$_2$ for which $\varepsilon_r$=3,9. We then obtain $C_{gc}$=1,9±1,5 aF and from Equation 5 we deduce an approximative number of transferred electrons of *1≤n≤8*. The large uncertainty comes from the cumulated uncertainty of each experimentally measured quantity, especially the charge retention hysteresis $\Delta V_h$. This interpretation in the framework of Coulomb blockade theory confirms the fact that this device involves a few electrons in each charge transfer, contrary to the thousands of electrons trapped in a conventional flash memory. Considering that each step has a constant height and width in the semi-log plot (Figure 5), the amount of charge transferred is constant which most probably is equal to a single electron.

## 4. Conclusion

In conclusion, this experiment provides a successful integration on a wafer scale of self assembled single electron memories working at helium temperatures. The double self-assembly process allows the controlled integration of both carbon nanotubes and gold floating dots on the device. Analysis of the temperature behaviour of the charge steps hysteresis have shown that single electron transfer events to and from the gold bead are well observed.

A better control of the gold bead position would reduce the dispersion of the island-electrometer coupling values. Moreover, reducing the size and thus the charging energy of the particle would increase the operation temperature which could reach 300 K. At low temperatures, the particle is coupled with the nearest quantum dot in the nanotube. Even though the single electron effect can subsist at 300K in nanotubes exhibiting dense defects [28], electron transport in CNFET is mostly controlled at room temperature by the Schottky barriers at the contacts. Placing the nanoparticle close to it, for example by controlled hydrophobic interactions [29] that allow specific attachment on the nanotube channel could be used to realize a similar device.

# Experimental section

**Substrate preparation**

We start from a 2 inch diameter degenerately doped silicon wafer. The wafer is thermally oxidized to obtain a 1-µm-thick silica top layer, which acts as the backgate dielectric. Sub-micron metallic electrodes are prefabricated using standard deep-UV lithography followed by electron beam evaporation and lift-off. They are made of 30 nm thick titanium layer covered by a thin film of cobalt (3 nm thick) that serves as a catalyst template for the nanotube chemical vapor deposition (CVD). The electrodes are separated by a 0.6 to 1.2 µm width gap.

**CNFET fabrication**

CVD deposition parameters were a 5-20 vol. % CH4 proportion in hydrogen, a 750-850°C substrate and a 30-100 mbar total pressure. The tungsten filament, placed 1 cm above the substrate, was heated up to 1900-2100°C. During the synthesis, in situ reflectivity measurements and the intensity of elastically scattered and reflected light from a 633 nm laser gave insight into real time growth kinetics.

**CNFET functionalization**

For the silanization of the substrate, we dip the wafer in (3-Aminopropyl)triethoxysilane (Sigma-Aldrich) diluted at 1% in desionized water during 1 hour. This provides a few nanometer thick amine coating on the whole sample, which is thick enough to embed the nanotubes. For gold nanoparticle attachment on the CNFET, we use an aqueous solution of colloidal gold (Ted Pella, Inc., Redding, Calif.) which is diluted in 10 vol. of water. A film of colloidald gold solution is deposited on half of the CNFET wafer and incubated for 5 min. The wafer is then gently rinsed in ultra-pure water. The other half of the wafer is left free from colloidal gold in order to provide control samples.

Encapsulation is performed with a film of parylene C deposited by cold CVD in a Labtop Model 3000 from Para-Tech. Finally top gate electrodes aligned above the nanotubes are fabricated by UV lithography followed by a thermal evaporation of 30 nm gold followed by lift-off.

**Electrical characterization**

The whole wafer was placed in a cryogenic probe station (TT-prober, Desert Cryogenics, Tucson, Az) that allows testing a several devices while the temperature can be continuously controlled between 2 K and 300 K. Very low temperature measurements were carried out in a He3/He4 dilution cryostat (Concept-Soudure, Echirolles, France), with shielded lines.